\documentclass[aps,prl,twocolumn]{revtex4}
\usepackage{graphicx}% Include figure files
\usepackage{dcolumn}% Align table columns on decimal point
\usepackage{bm}% bold math

\begin{document}
 
\title{Suppression of Bragg scattering by collective interference of 
spatially ordered atoms with a high-Q cavity mode}
 
\author{Stefano Zippilli$^1$}
\author{Giovanna Morigi$^1$}\author{Helmut Ritsch$^2$}
\affiliation{$^1$ Abteilung f\"ur Quantenphysik,
Universit\"at Ulm, Albert-Einstein-Allee 11, D-89069 Ulm, Germany.\\
$^2$ Institut f\"ur Theoretische Physik,
Universit\"at Innsbruck, Technikerstrasse 25/2, A-6020 Innsbruck, Austria}
\date{\today}
\begin{abstract}                       
When N driven atoms emit in phase into a high-Q cavity mode, 
the intracavity field generated by collective scattering
interferes destructively with the pump driving the atoms. Hence atomic
fluorescence is suppressed and cavity loss becomes the dominant decay channel
for the whole ensemble. 
Microscopically 3D light-intensity minima are formed in the vicinity of the
atoms that prevent atomic excitation and form a regular
lattice. The effect gets more pronounced for large atom numbers,
when the sum of the atomic decay rates exceeds the rate of cavity losses and one would
expect the opposite behaviour. These results provide new insight into recent
experiments on collective atomic dynamics in cavities.
\end{abstract}

\maketitle

Recent experimental progress has allowed to trap cold atomic gases within
high-$Q$ optical resonators. In ring resonators collective oscillations 
of the atoms has been reported~\cite{Hemmerich03} and strong accelerations were 
demonstrated~\cite{Zimmermann03}. In pioneering experiments with transversally
pumped atoms in a standing-wave cavity very efficient cooling of large ensembles was achieved
by Vuletic and coworkers~\cite{Chan03,Black03}. They found strong coherent emission into the cavity mode,
which exceeded the rate of scattering into free space by orders of
magnitude~\cite{Chan03,Black03}. 
The cavity field randomly attained one of two well defined phases with
$\pi$-difference. This behaviour gives strong evidence of formation of a regular atomic pattern with wavelength
periodicity, so that the atoms scatter in phase into the cavity mode, while
the observed $\pi$-phase jumps correspond to transitions to an alternative pattern shifted by half wavelength.
This hypothesis is supported by numerical results~\cite{Domokos02}, showing self-organization into the pattern in the parameter regime of~\cite{Chan03,Black03}. In this context, 
the enhanced rate of cavity emission has been interpreted as a signature of
Bragg scattering
. However, Bragg scattering by itself does not explain the dramatic
suppression of atomic fluorescence, which makes cavity decay the
dominant channel of dissipation.

In this Letter we argue that the origin of the enhanced cavity emission rate
and the suppression of atomic fluorescence can be traced back to quantum
interference of coupled oscillators, namely the field mode and the collective
atomic polarization. 
This model reproduces qualitatively several features of the 
stationary dynamics observed in~\cite{Black03}, even though in the experiment the
atoms had a more complex internal structure and more than a single mode was
involved. It predicts that, when the atoms are organized in a regular
spatial pattern emitting in phase into the cavity mode,
stationary cavity field and pump have opposite
phases and mutually cancel at the atomic sites.
Thus, the atoms are not excited and fluorescence as well as superradiant
scattering into the cavity mode are suppressed. These dynamics are found in
high-Q resonators, and differ radically from the dynamics at the basis of
cavity-enhanced emission 
observed in experiments with a similar setup but in the
bad-cavity limit regime~\cite{Heinzen}. As in optical
bistability~\cite{OpticalBistability} the effect occurs in the strong coupling
regime, but it is characterized by only one steady state, for which the atoms
are in the ground state. In this case the amplitude of the cavity field
cancelling the pump field is independent of the number of atoms. Hence, 
with properly rescaled parameters the effect can be also found for a single
atom. For many atoms efficient suppression of excitation is
only present when they emit in phase into the cavity mode. The  
suppression of fluorescence is accompanied by a coherent cavity field, which 
gives a clear signature of pattern formation. This interpretation is supported 
by an analysis showing that the regular spacing of the atoms is a stable
configuration in the parameter regime of~\cite{Black03}.  

\begin{figure}[h]
\includegraphics[width=5cm]{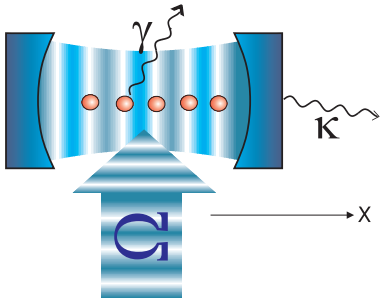}
\caption{N atoms are inside a 1D optical resonator and homogeneously driven
by a laser propagating in the transverse direction. The atoms are localized 
according to a spatial pattern such that they emit in phase into the cavity 
mode
(see~\protect\cite{Domokos02}).}
\label{Fig1}
\end{figure}

We consider a single standing-wave cavity mode resonantly coupled to $N$
atomic dipoles with spatially 
dependent coupling constant $g=g_0\cos (2\pi
x/\lambda)$, where $\lambda$ is the mode wavelength and $x$ gives the
position along the cavity axis. In addition, the dipoles couple with Rabi
frequency $\Omega$ to a plane-wave field at frequency $\omega_L$ and
propagating orthogonal to the cavity axis, as illustrated in Fig.~\ref{Fig1}. Here we assume that the
atoms are point-like and distributed at the positions $x_n^{(0)}=x+n\lambda$,
where $n$ is an integer such that $g(x_n^{(0)})=g\neq 0$. This assumption 
will be later justified by showing that this is in fact a mechanically stable
situation. The pattern is assumed to have a low filling factor,
such that collective radiative effects in free space are negligible.  
The coherent dynamics of the system is described by
the Hamiltonian $H=-\hbar\delta_ca^\dagger a+\sum_{n=1}^N H_n$, with $a$,
$a^\dagger$ annihilation and creation operators of a cavity photon, and
\begin{eqnarray} \label{H} H_n=\hbar\Delta \sigma_n\sigma_n^{\dagger}+
  \hbar[\sigma_n^{\dagger}(g(x_n)a+\Omega)+{\rm H.c.}] \end{eqnarray} with
$\sigma_n$, $\sigma_n^{\dagger}$ dipole operators for the $n$-th atom, and
$\delta_c=\omega_{\rm L}-\omega_c$, $\Delta=\omega_{\rm L}-\omega_0$,
detunings of the laser from the frequency of the cavity and of the dipole,
respectively. The master equation for the density matrix $\rho$ of atoms and
cavity mode is $\frac{\partial}{\partial t}\rho =[H,\rho]/{\rm i}\hbar+{\cal
  L}\rho  + {\cal K}\rho$, where the superoperator ${\cal L}$ describes
damping due to spontaneous decay at rate $\gamma$, while ${\cal K}\rho$
describes cavity decay with zero-photon linewidth
$\kappa$~\cite{Carmichael}. Note that for the moment we consider a
one-dimensional model and neglect the center-of-mass motion.

\begin{figure}[h]
\includegraphics[width=9cm]{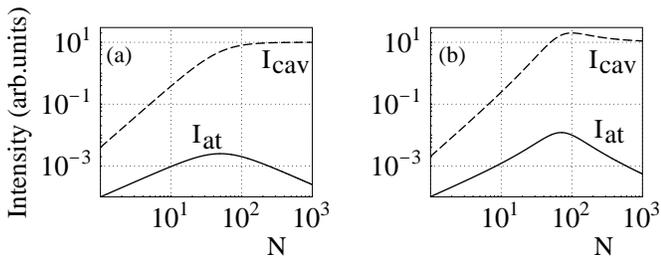}
\caption{Intensity of the signal at the cavity mirror ($I_{\rm cav}$, dashed
  line) and total atomic fluorescence intensity ($I_{\rm at}$, solid line) as
  a function of $N$ for
$\Omega=g=10\gamma$, $\kappa=10\gamma$, $\Delta=-1000\gamma$, and (a)
$\delta_c=0$, (b) $\delta_c=-5\gamma$. }
\label{Fig2}
\end{figure}

In the parameter regime when the collective atomic dipole is driven below
saturation, for $|\gamma/2+i\Delta|\gg \sqrt{N}g,\sqrt{N}\Omega$, the
stationary field amplitude takes the form \begin{equation} \label{beta:max}
  \alpha=-\frac{\Omega}{g_0} \frac{Ns(\gamma/2+{\rm
      i}\Delta)}{Ns(\gamma/2+{\rm i}\Delta)+\kappa/2-{\rm i}\delta_c}
\end{equation} and the occupation of the excited state of each atom is
\begin{equation} \label{Excited}
  \Pi_{n}=\frac{\Omega^2}{(\gamma/2)^2+\Delta^2}
  \frac{\kappa^2/4+\delta_c^2}{(Ns\gamma+\kappa)^2/4+(Ns\Delta-\delta_c)^2}
\end{equation} where $s=g^2/(\Delta^2+\gamma^2/4)$. In
  Eqs.~(\ref{beta:max}-\ref{Excited}) 
the number of atoms $N$ scales the terms containing the atomic
parameters. In particular, for $N$ sufficiently large the occupation of the atomic excited state vanishes as
$1/N^2$ while the field inside the
cavity tends to the constant value $\alpha_0= -\Omega/g$, which neither
depends on the detunings $\Delta$, $\delta_c$, nor on the cavity and atomic
decay rates. This behaviour is visible in Fig.~\ref{Fig2}, where the total fluorescence
intensity $I_{\rm at}=N\gamma\Pi_n$ and the signal at the cavity mirror
$I_{\rm cav}=\kappa |\alpha|^2$ are plotted as a function of $N$.
Here a threshold value $N_0$ can be identified that separates two different
dynamics, corresponding to the regimes of weak and strong coupling. In fact, 
for $N\ll N_0$ the excited state population in (\ref{Excited}) 
is approximately given by the value in free space, and the field intensity
scales quadratically with the number of atoms: There is no back--action of the
cavity on the atomic dynamics, since the cavity decay rate is faster than the
rate at which the atomic degrees of freedom reach their steady state. This
is the regime where Bragg enhancement of superradiant scattering into the
cavity mode is found. For $N\gg N_0$,
on the other hand, the total power dissipated by spontaneous emission scales
with $1/N$. Thus the system dissipates mainly through cavity loss, where the
signal at the cavity output is constant and, remarkably, independent of $N$.
Hence, for $N\gg N_0$ there is no signature of Bragg scattering. 
At $\Delta=0$ this regime corresponds to a large 
cooperativity parameter $C=Ng_0^2/\gamma\kappa >1$. 
A situation closely related to the experimental parameters of~\cite{Black03}
is found for the case $|\Delta|\gg \gamma,\kappa$ and $\delta_c\neq 0$,
illustrated in Fig.~\ref{Fig2}(b). Here the
critical number of atoms $N_{0}\sim |\delta_c\Delta|/g_0^2$ denotes the
situation when the laser drives resonantly the collective resonance of
the atoms and cavity system, manifesting itself in the enhancement of the two
signals visible in Fig.~\ref{Fig2}(b). When the limit $N\gg N_0$ is
reached, on the other hand, the system behaves like in the resonant case,
namely the cavity field is independent of the number of
atoms and the atoms do not fluoresce.
\begin{figure}[h]
\includegraphics[width=7cm]{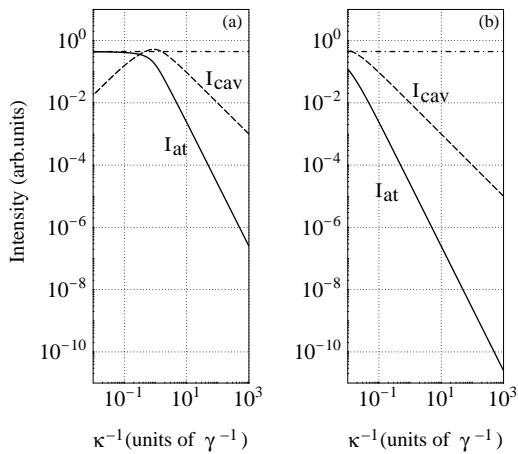}
\caption{Intensity of the signal at the cavity mirror ($I_{\rm cav}$, dashed
  line) and of the atomic fluorescence signal ($I_{\rm at}$, solid line)
as a function of $\kappa^{-1}$ for a single atom. 
Here $\Omega=\gamma$, $\Delta=\delta_c=0$ and (a) $g=\gamma$, (b)
$g=10\gamma$. The dashed-dotted line corresponds to the free space fluorescence rate.}
\label{Fig3}
\end{figure}
\begin{figure}[h]
\includegraphics[width=9cm]{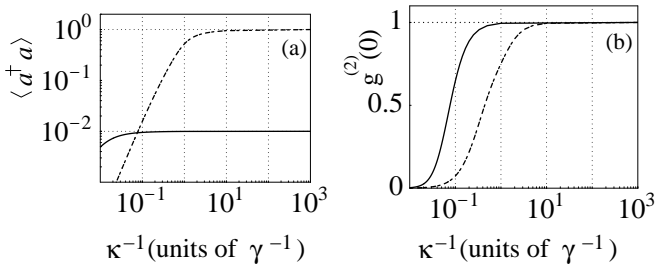}
\caption{(a) Mean number of cavity photon and (b)
second-order correlation function $g^{(2)}(0)$ as a function of
$\kappa^{-1}$ for $\Delta=\delta_c=0$, $\Omega=\gamma$ and $g=\gamma$
(dashed line), $g=10\gamma$ (solid line).}
\label{Fig4}
\end{figure}

These results can be understood in the limiting case
$\kappa=0$ and $\delta_c=0$. For these values the dependence on $N$ drops out
from Eqs.~(\ref{beta:max}-\ref{Excited}) and one obtains $\alpha=\alpha_0$
and $\Pi_n=0$. In this case the dynamics are scalable down to 
a single atom and are exactly solvable~\cite{Alsing92}. Here
the steady state $\rho_{\rm st}$ is a pure state, with the
cavity field in the coherent state at the amplitude $\alpha_0$ and the atom in
the ground state. Thus, no fluorescence photons are emitted. This behaviour
is due to
destructive interference between pump and cavity field that drive
the atom with equal intensity but opposite phase, as one can verify by
applying the Hamiltonian~(\ref{H}) to the system prepared in $\rho_{\rm st}$.
These dynamics are
independent of the intensity of the pump, which only determines the mean
number of cavity photons at steady state through the ratio
$\Omega/g$~\cite{Footnote}. 

Several features of these dynamics largely survive
for finite and fairly large values of the cavity damping $\kappa$. In
particular, the signal transmitted at the cavity mirror $I_{\rm cav}=\kappa
{\rm Tr}\{a^{\dagger}a\rho\}$ can be orders of magnitude larger than the
fluorescence signal $I_{\rm at}=\gamma{\rm Tr}\{\sigma^{\dagger}\sigma
\rho\}$, as illustrated in Fig.~\ref{Fig3}. In the limits $\Delta=0$ and
$g\gg\gamma\gg\kappa$, these signals have the form $$I_{\rm at}\approx \kappa
|\alpha_0|^2/8C_1;~~~~I_{\rm cav}\approx
\kappa|\alpha_0|^2\left(1-1/8C_1\right)$$ where $C_1=g^2/2\gamma\kappa$ is the
cooperativity parameter per atom~\cite{Kimble94}. Hence,
the regime of small $\kappa$ in Fig.~\ref{Fig3} corresponds to a large
cooperativity parameter, like the regime
of large $N$ in Fig.~\ref{Fig2}. For $C_1\gg 1$ the field
inside the cavity is still well approximated by a coherent state, as it can be
verified 
from the second-order correlation function $g^{(2)}(0)$ at the cavity
mirror. In Fig.~\ref{Fig4}(b) $g^{(2)}(0)$ is plotted as a function of
$\kappa^{-1}$, showing that the cavity field exhibits a Poissonian behaviour for a
fairly large range of values of $\kappa$ and even for a very small number of
photons inside the cavity (solid line in Fig.~\ref{Fig4}). Thus the cavity
mode is in a coherent state, independently of the mean energy of the cavity
field. This behaviour contrasts dramatically with the antibunching observed
when the pump is set directly on the cavity~\cite{Kimble94,Brecha99}.
 
Further insight is gained from the rate of photon scattering obtained by
scanning an additional weak transverse pump across atomic resonance. 
Denoting with $\delta_P=\omega_P-\omega_c$ the
 detuning of the probe from the cavity frequency, for $\delta_c=0$ and
 $\kappa=0$ the atom scatters photons at the rate \begin{equation}
   w(\delta_P)\propto \frac{\gamma\delta_P^2}
   {[\delta_P(\delta_P+\Delta)-g(x)^2]^2+\delta_P^2\gamma^2/4}
   \label{Excitation:Rate}\end{equation}  
which vanishes at $\delta_P=0$, thereby exhibiting a
Fano--like profile at this point~\cite{Lounis92,Rice96}. 
It is remarkable that the pump intensity
$\Omega$, and thus the mean number of cavity photon, does not appear in 
Eq.~(\ref{Excitation:Rate}). In particular, the positions of the two maxima of
$w(\delta_P)$ correspond to the energies of the dressed states of the atom
in an {\it empty} cavity. This behaviour, which is independent of the cavity
 field energy, is due to the vanishing electric field at the atomic
 position. 

In the limit $\kappa=0$ and $\delta_c=0$ these results can be scaled with the
number $N$ of atoms. Then, the Hamiltonian $H$ describes the
 Jaynes-Cunning dynamics of a collective dipole coupling to the cavity mode
 with $g\to\sqrt{N}g$, provided that the atoms are localized and distributed
 according to the spatial pattern. At steady state, when the
collective dipole is driven below saturation,  the atoms are in the ground
state and the cavity field is a coherent state of amplitude
$\alpha_0$~\cite{FootnoteDicke}, while the splitting
of the maxima in $w(\delta_P)$ scales according to the rule $g\to\sqrt{N}g$.
Note that the scaling can be applied only in this ideal case, where one has
always strong coupling. For bad cavities and $\delta_c\neq 0$, 
on the other hand, only a large number of atoms allows to achieve
the necessary large cooperativity for accessing these dynamics.
In contrast to optical bistability, however, here 
no bistable behaviour is found as the atoms are in the ground state.

\begin{figure}[h]
\includegraphics[width=9cm]{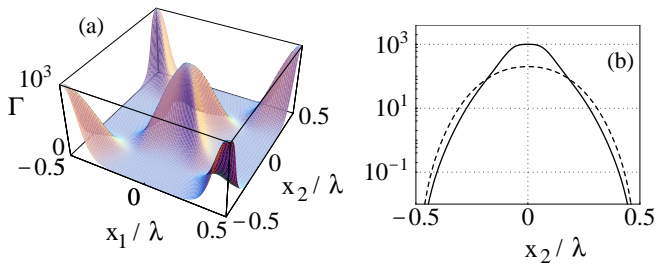}
\caption{(a) Ratio $\Gamma=I_{\rm cav}/I_{\rm at}$ for two atoms as a function
  of their position $x_1$ and $x_2$ 
inside the cavity.  Here,
$g_0=10\gamma$, $\Delta=100\gamma$, $\delta_c=0$, $\Omega=\gamma$ for both
atoms, and $\kappa=0.2\gamma$. (b) Ratio as in
(a) as a function of $x_2$ for $x_1=0$. The dashed line shows $\Gamma$ for
the same parameters but $\kappa=\gamma$. Note that $x_1,~x_2$ are
plotted modulus $\lambda$, and $x_1\neq x_2$.}   
\label{Fig5}
\end{figure}

In Fig.~\ref{Fig5} the ratio $I_{\rm cav}/I_{\rm at}$ is displayed 
for two atoms at the positions $x_1$ and $x_2$, showing clearly
that this ratio is maximum when the
atoms are a wavelength apart. For $\kappa\neq 0$ 
absolute maxima are found when the
atoms are at the antinodes of the cavity mode,  where the cooperativity
parameter is largest and the total electric field practically vanish. 
The three dimensional pattern is found 
taking into account the phase of the pump. Then, 
the zeros of the electric field
are distributed according to a 
Body-Centered-Cubic lattice with distance $\lambda/2$ between adiacent
planes~\cite{Domokos02,Black03}. Fluorescence is suppressed when the atoms are
localized at these points, thus forming a stationary pattern.
This latter condition constitutes a substantial
difference to the collective scattering via acceleration observed in the
dynamics of the collective
atomic recoil laser~\cite{Zimmermann03,CARL}. 

We study now the mechanical stability 
of the atomic pattern, which is a central ingredient 
in our model. We use the coupled semiclassical equations for field and atomic motion~\cite{Domokos02}
to describe the parameter regime of~\cite{Black03}, where the atoms are driven 
at far-off resonance and the relative fluctuations of
the cavity-field amplitude are small, and assume $N\gg N_0$, so that at leading
order in $N$ the cavity field amplitude is $\alpha_0$. A trivial equilibrium
configuration is found when the atoms are at the antinodes of the cavity mode
with spacing equal to the wavelength $\lambda$. Stability for
small fluctuations $\delta x_n$ of the atomic positions is found when the
first derivative of the semiclassical force at these points is negative, which
corresponds to the
condition~\cite{Footnote2}\begin{equation}\label{Mechanical}\delta f_n\sim
  2\hbar  k^2(\Omega/g_0)^2\delta_c/N<0\end{equation} that is 
$\delta_c<0$. Notably the stability condition 
is not affected by the sign of the detuning $\Delta$. 
This behaviour has been verified numerically. At these points 
the intensity of the force $\delta f_n\delta
x_n$ is proportional to the average number of cavity photons and thus to the pump intensity. Hence, 
the mechanical potential at these positions gets steeper for
larger pumps. This
dependence is consistent with the threshold behaviour 
measured in~\cite{Chan03,Black03}. Also the dependence on the detunings
is in line with the experimental observations, that found enhanced cavity
emission for both signs of the detuning $\Delta$ but for a finite range of
values of $\delta_c<0$~\cite{Chan03,Black03}. 
It is an open and intriguing question 
how the system evolves into the self-organized pattern and how the
non-equilibrium dynamics depend on the various parameters. Moreover, do other
(meta)-stable equilibrium states exist? And how does noise affect the pattern
stability?  Such questions
can be addressed using the theory developed in~\cite{HelmutReview}. 

For systems of one or few atoms in high-Q cavities this interference effect
has numerous potential applications. As the atoms generate a position dependent
field without being excited, this could be a new version of nondestructive
single atom detection. In particular molecules with no closed cycle could be efficiently detected.
Further possibilities are implementations of conditional coherent dynamics as
needed for quantum information processing. Here one expects
less decoherence as the atoms interact while being almost in the ground state. 
Scattering from internal atomic superposition states would immediately create entanglement even in the steady
state. Moreover, the coherence properties of the transmitted signal, which are
preserved even for very small photon numbers, suggest an alternative kind of
photon-emitters to the one investigated in~\cite{Kuhn02,Kimble03}. These
investigations may be realized with present experimental setups, that can trap
single or few atoms and couple them to the cavity
field in a controlled way~\cite{Eschner01,MPQ01,Mundt02,Sauer03,KimbleFORT03}. 
The authors gratefully ackowledge discussions with H.J.\ Carmichael, A.\ Kuhn,
H.\ Mabuchi, G.\ Rempe, W.\ Schleich. This work has been supported by the
EU-networks QUEST and QGATES.


\begin{thebibliography}{99}
\bibitem{Hemmerich03}
%Collective atomic motion in an optical lattice formed inside a high-Q cavity
%B.\ Nagorny, Th.\ Els\"asser, A.\ Hemmerich,
B.\ Nagorny {\it et al}, Phys.\ Rev.\ Lett.\ {\bf 91},
153003 (2003).
 
\bibitem{Zimmermann03}
%Observation of lasing mediated by collective atomic recoil
%Kruse, C.\ von Cube, C.\ Zimmermann, Ph.W.\ Courteille
D.\ Kruse {\it et al}, Phys.\ Rev.\ Lett.\
{\bf 91}, 183601 (2003).

\bibitem{Chan03}
%Observation of collective emission induced cooling of atoms in an optical
%cavity H.W.\ Chan, A.T.\ Black, V.\ Vuletic,
H.W.\ Chan {\it et al}, Phys.\ Rev.\ Lett.\ {\bf 90}, 063003
(2003).

\bibitem{Black03} 
%Observation of collective friction forces due to spatial self-organization of
%atoms: from Rayleigh to Bragg scattering
%A.T.\ Black, H.W.\ Chan, V.\ Vuletic
A.T.\ Black {\it et al}, Phys.\ Rev.\ Lett.\ {\bf 91}, 203001
(2003).
  
\bibitem{Domokos02}
%Collective cooling and self-organization of atoms in a cavity
P.\ Domokos and H.\ Ritsch, Phys.\ Rev.\ Lett.\ {\bf 89}, 253003 (2002).

\bibitem{Heinzen} 
D.J.~Heinzen {\it et al}, Phys.\ Rev.\ Lett.\ {\bf 58}, 1320 (1987); 
D.J.~Heinzen and M.S.~Feld, {\it ibid.} {\bf 59}, 2623 (1987). 

\bibitem{OpticalBistability}
R.\ Bonifacio and L.A.\ Lugiato, Opt.\ Comm.\ {\bf 19}, 172 (1976);
Phys.\ Rev.\ A {\bf 18}, 1129 (1978);
L.A.\ Lugiato in {\it Progress in Optics} vol.\ XXI, pp.\ 69ff,
ed.\ E.\ Wolf (North-Holland, Amsterdam 1984).
 
\bibitem{Carmichael}
H.J.\ Carmichael, {\it An Open Systems Approach to Quantum Optics},
Springer-Verlag (Berlin, Heidelberg, New York, 1993).
 
\bibitem{Alsing92}
%suppression of fluorescence in a lossless cavity
%P.M.\ Alsing, D.A.\ Cardimona, H.J.\ Carmichael,
P.M.\ Alsing {\it et al}, Phys.\ Rev.\ A {\bf 45}, 1793
(1992).
 
\bibitem{Footnote}
Alternatively, for fixed pump
intensity the mean energy of the field 
depends on the atomic position through $g(x)$. 

\bibitem{Kimble94}
H.J.\ Kimble, in {\it Cavity Quantum Electrodynamics}, p. 203, ed.\ by P.R.\
Berman, Academic Press (New York, 1994). 


\bibitem{Brecha99}
R.\ Brecha {\it et al}, Phys.\ Rev.\ A {\bf 59}, 2392 (1999).

\bibitem{Lounis92}
B.\ Lounis and C.\ Cohen-Tannoudij, J.\ de Phys.\ II (France) {\bf 2}, 579
(1992).

\bibitem{Rice96}
%cavity induced transparency
P.R.\ Rice and R.J.\ Brecha, Opt.\ Comm.\ {\bf 126}, 230 (1996).

\bibitem{FootnoteDicke}
At saturation and for large filling factors dipole-dipole interaction
may modify substantially the dynamics. See for example
J.P.~Clemens {\it et al}, Phys.\ Rev.\ A {\bf 68}, 023809 (2003).  

\bibitem{CARL}
R.\ Bonifacio and L.\ DeSalvo, Nucl.\ Instrum.\ Methods {\bf 341}, 360 (1994); 
R.\ Bonifacio {\it al}, Phys.\ Rev.\ A {\bf 50}, 1716 (1994).
  
\bibitem{Footnote2}
This condition is
sufficient, since the field amplitude $\alpha$ does not vary in first order in
$\delta x_n$, nor does the force for small fluctuations in $\alpha$.

\bibitem{HelmutReview}
P.\ Domokos and H.\ Ritch, J.\ Opt.\ Soc.\ Am.\ B {\bf 20}, 1098 (2003).
 
\bibitem{Kuhn02}
%A.\ Kuhn, M.\ Hennrich, and G.\ Rempe,
A.\ Kuhn {\it et al}, Phys.\ Rev.\ Lett.\ {\bf 89}, 067901
(2002).
 
\bibitem{Kimble03}
%Experimental realization of a one-atom laser in the regime of strong coupling
%J.\ McKeever, A.\ Boca, A.D.\ Boozer, J.R.\ Buck, H.J.\ Kimble, Nature
J.\ McKeever {\it et al}, Nature (London) {\bf 425}, 268 (2003).
 
 \bibitem{Sauer03}
%Cavity QED with optically transported atoms
%J.A.\ Sauer, K.M.\ Fortier, M.S.\ Chang, C.D.\ Hamley, M.S.\ Chapman
J.A.\ Sauer {\it et al}, Phys.\ Rev.\ A {\bf 69}, 051804 (2004).
 
\bibitem{KimbleFORT03}
%State-insensitive cooling and trapping of single atoms in an optical cavity
%J.R.\ Buck, A.D.\ Boozer, A.\ Kuzmich, H.C.\ N\"agerl, D.M.\
%Stamper-Kurn, H.J.\ Kimble
J.\ McKeever {\it et al},
Phys.\ Rev.\ Lett.\ {\bf 90}, 133602 (2003).
 
 
\bibitem{Eschner01}
%J.\ Eschner, Ch.\ Raab, F.\ Schmidt-Kaler, and R.\ Blatt
J.\ Eschner {\it et al}, Nature {\bf 413}, 495
(2001).
 
 
\bibitem{MPQ01}
%G.R.\ Guth\"ohrlein, M.\ Keller, K.\ Hayasaka, W.\ Lange, and H.\ Walther,
G.R.\ Guth\"ohrlein {\it et al}, Nature {\bf 414}, 49 (2001).
 
\bibitem{Mundt02}
%A.\ Kreuter, C.\ Becher, D.\ Leibfried, J.\ Eschner, F.\ Schmidt-Kaler, and R.\ Blatt
A.B.\ Mundt {\it et al}, Phys.\ Rev.\ Lett.\ {\bf 89}, 103001 (2002).
 
 
\end{thebibliography}
\end{document}